\documentclass{article} 
  \usepackage{cite}
  \usepackage[pdftex]{graphicx}
\usepackage{amsmath}
 \usepackage[caption=false,font=footnotesize,labelfont=sf,textfont=sf]{subfig}
\usepackage{url}

\usepackage{booktabs}
\usepackage{multirow}
\usepackage[super]{nth}
\usepackage{soul} 
\usepackage{color}
\usepackage{authblk} 

\usepackage{etoolbox}
\apptocmd{\thebibliography}{\setlength{\itemsep}{0pt}}{}{}

\begin{document}
%
\title{A method to evaluate the reliability of social media data for social network analysis}

\author[1,2]{Derek Weber}
\author[3-5]{Mehwish Nasim}
\author[5,6]{Lewis Mitchell}
\author[2,7]{Lucia Falzon}
\affil[1]{School of Computer Science, University of Adelaide, Australia}
\affil[2]{Defence Science and Technology Group, Adelaide, Australia}
\affil[3]{Data61, Commonwealth Scientific and Industrial Research Organisation, Adelaide, Australia}
\affil[4]{Cyber Security Cooperative Research Centre, Adelaide, Australia}
\affil[5]{ARC Centre of Excellence for Mathematical and Statistical Frontiers, Adelaide, Australia}
\affil[6]{School of Mathematical Sciences, University of Adelaide, Australia}
\affil[7]{School of Psychological Sciences, University of Melbourne, Australia}


%


\maketitle

\begin{abstract}
To study the effects of Online Social Network (OSN) activity on real-world offline events, researchers need access to OSN data,
the reliability of which has particular implications for social network analysis. 
This relates not only to the completeness of any collected dataset, but also to constructing meaningful social and information networks from them. 
In this multidisciplinary study, we consider the question of constructing traditional social networks from OSN data and then present a measurement case study showing how the reliability of OSN data affects social network analyses.
To this end we developed a systematic comparison methodology, which we applied to two parallel datasets we collected from Twitter.
We found considerable differences in datasets collected with different tools and that these variations significantly alter the results of subsequent analyses.

Our results lead to a set of guidelines for researchers planning to collect online data streams to infer social networks.
\end{abstract}


%

\section{Introduction}

Online activities can be associated with dramatic offline effects, such as COVID19 misinformation leading to panic buying of toilet paper\footnote{https://theconversation.com/coronavirus-why-people-are-panic-\newline buying-loo-roll-and-how-to-stop-it-133115}
, online narratives incorrectly attributing Australia's ``Black Summer'' bushfires to arson amplifying attention to it in the media~\cite{WeberNFM2020arson}, and attempts to influence domestic and foreign politics~\cite{ratkiewicz2011,woolley2018us}
. For researchers to successfully analyse online activity and provide advice about protection from such events, they must be able to reliably analyse data from online social networks (OSNs).

Social Network Analysis (SNA) facilitates exploration of social behaviours and processes. 
OSNs are often considered convenient proxies for offline social networks, because they seem to offer a wide range of data on a broad spectrum of individuals, their expressed opinions and inter-relationships. 
It is assumed that the social networks present on OSNs 
can inform the study of information dissemination and opinion formation, contributing to an understanding of offline community attitudes.
Though such claims are prevalent in the social media literature, there are serious questions about their validity due to an absence of SNA theory on online behaviour, the mapping between online and offline phenomena, and the repeatability of such studies. 
In particular, the issue of reliable data collection is fundamental. 
Collection of OSN data is often prone to inaccurate boundary specifications due to sampling issues, collection methodology choices, as well as platform  constraints. 

Previous work has considered the question of data reliability from a sampling perspective~\cite{morstatter2013sample,gonzalez2014assessing,JosephLC2014comparison,PaikLin2015}, biases~\cite{RuthsPfeffer2014,tromble2017we,pfeffer2018tampering,OlteanuCDK2019} and the danger of making invalid generalisations using ``big data'' approaches lacking nuanced interpretation of the data~\cite{lazer2014parable,tufekci2014big}. 
Analyses of incomplete networks exist~\cite{HolzmannAK2018}, but 
this paper specifically considers the questions of data reliability for SNA, 
considering not only the significance of 
online 
interactions to discover meaningful social networks, but also how sampling and boundary issues can complicate analyses of the networks constructed. Through an exploration of modelling and collection issues, and a 
measurement study examining the reliability of simultaneously collected, or \emph{parallel}, datasets, this multidisciplinary study addresses the following research question:
\begin{quote}
    \textit{How do variations in collections affect the results of social network analyses?}
\end{quote}
Our work makes the following contributions:

\begin{enumerate}
    \item Discussion of the challenges mapping OSN data to meaningful social and information networks;
    \item A methodology for systematic dataset comparison; 
    \item Recommendations for the use and evaluation of social media collection tools; and
    \item Two original social media datasets collected in parallel, and relevant analysis code\footnote{
    \url{https://github.com/weberdc/socmed_sna}}.
\end{enumerate}


This paper continues with four sections: 1) challenges obtaining and modelling social networks from OSN data for SNA; 2) systematic parallel dataset comparison methodology; 3) results from using our methodology in a two-phase case study; and 4) recommendations for social media researchers and analysts, and directions for future research.







\section{Social networks from social media} 

Using SNA to explore social behaviours and processes from OSN data presents many challenges. Most easily accessible OSN data consists of timestamped interactions, rather than details of long-standing relationships, which form the basis of SNA theory. Additionally, although interactions on different OSNs are superficially similar
, how they are implemented 
may subtly alter their interpretation. They offer a window onto online behaviour only, and any implications for offline relations and behaviour are unclear. Beyond 
modelling and reasoning with the data is the question of 
collection -- accessing the right data to construct meaningful social networks is challenging. 
OSNs 
provide a limited subset of their data through a variety of mechanisms, balancing privacy and 
competitive advantage with openness and transparency. 



\subsection{Interactions and relationships online} \label{sec:online_interactions}

SNA provides concepts and tools to model social networks among actors and using network-based techniques to study social behaviours and processes~\cite{borgatti2013analyzing}. Given the availability, nature and structure of much OSN data, they offer unique opportunities to apply SNA. 
Relationships between social network actors must be stable, though they may be dynamic~\cite{wasserman1994social,nasim2016inferring,borgatti2009network}. 
On social media, accounts can easily fulfil the role of actors, but precisely what constitutes a relationship is unclear. 
A common semi-permanent candidate is the \emph{friend} or \emph{follower} relationship common to most OSNs, 
but, due to 
how OSNs present their specific features to users,
each online community develops its own interaction culture. 
Therefore, such connections do not necessarily easily translate between OSNs. 
Is a Facebook \emph{friend}ship really the same as a \emph{follow} on Twitter, even if reciprocated? And how do each relate to offline friendships? 

OSNs offer ways to establish and maintain relations with others. 
Specific interactions may be visible to different accounts, intentionally or incidentally (\emph{cf.} replying versus using a hashtag). Exploration of these differences may lead to an understanding of the author's intent and  the identity of the intended audience. 
Is replying to a politician's Facebook post a way to connect directly with the politician, or is it a way to engage with the rest of the community replying to the post, either by specifically engaging with dialogue or merely signalling one's presence with a comment of support or dismay? 
A reply could be all of these things but, in particular, it is evidence of engagement at a particular time and indicates information flow between individuals~\cite{bagrow2019information}. Since most interactions are directed they offer an opportunity to study the flow of information and influence. 
On the other hand, although friend and follower connections can indicate community membership, they obscure the currency of that connection. 
Through their dynamic interactions, a user who \emph{liked} a Star Wars page ten years ago can be distinguished from one who not only \emph{liked} it, but posted original content to it on a monthly basis.
Therefore, we specifically focus on interactions rather than membership relations in this study.


\subsection{Social Network Analysis theory}

Relationships between individuals in a social network may last for extended periods of time, vary in strength, and be based upon a variety of factors, not all of which are easily measurable. 
Because of the richness of the concept of social relationships, data collection is often a qualitative activity, involving directly surveying community members for their perceptions of their direct relations and then perhaps augmenting that data with observation data such as recorded interactions (e.g., meeting attendance, emails, phone calls). 
It is tempting to believe that this richness should be discoverable in the vast amount of interaction data available from OSNs, but there are issues to consider:
\begin{enumerate}
    \item links between social media accounts may vary in type and across OSNs --- it is unclear how they contribute to any particular relationship;
    \item what is observed online is only a partial record of interactions in a relationship, where interactions may occur via other OSNs or online media, or entirely offline; and
    \item collection strategies and OSN constraints may also hamper the ability to obtain a complete dataset.
\end{enumerate}
Although many interactions seem common across OSNs (e.g., a retweet on Twitter, a repost on Tumblr and a share on Facebook), nuances in how they are implemented and how data retrieved about them is modeled (beyond questions of semantics) may confound direct comparison. For example, a Twitter retweet refers directly to the original tweet, obscuring any chain of accounts through which it passed to the retweeter~\cite{RuthsPfeffer2014}. 
There are efforts to probabilistically regenerate such chains~\cite{DebateNightICWSM2018,gray2020bayesian}, but, in any case, is one account sharing the post further evidence of a relationship? What if it is reciprocated once, or three times? What if the reciprocation occurs only over some interval of time? These questions require careful consideration before SNA can be applied to OSN data. 

\subsection{Challenges obtaining OSN data}

Social media data is typically accessed via an OSN's Application Programming Interfaces (APIs), which place constraints on how true a picture researchers can form of any relationship. Via its API an OSN can control: \emph{how much data} is available, through rate limiting, biased or at least non-transparent sampling, and temporal constraints; \emph{what types of data} are available, through its data model; and \emph{how precisely data can be specified}, through its query syntax. 
Many OSNs offer commercial access, which provides more extensive access for a price, though use of such services in research raises questions of repeatability~\cite{RuthsPfeffer2014}. 
This is done to protect users' privacy but also to maintain competitive advantage. 
Researchers must often rely on the cost-free APIs, which present further issues. 
Twitter's 1\% Sample API 
has been found to provide highly similar samples to different clients and it is therefore unclear whether these are truly representative of Twitter traffic~\cite{JosephLC2014comparison,PaikLin2015}. 
Studying social media data therefore raises questions about the ``the coverage and representativeness''~\cite[p.17]{gonzalez2014assessing} of the sample obtained and how it therefore ``affects the networks of communication that can be reconstructed from the messages sampled''~\cite[p.17]{gonzalez2014assessing}. 

Empirical studies have compared the inconsistencies between collecting data from search and streaming APIs using the same or different lists of hashtags. 
Differences have been discovered between the free streaming API and the full (commercial) ``firehose'' API~\cite{morstatter2013sample}.
There is general agreement in the literature that the consistency of networks inferred from two streaming samples is greater when there is a high volume of tweets even when the list of hashtags is different~\cite{gonzalez2014assessing}.
More concerning is the ability to tamper with Twitter's sample API to insert messages~\cite{pfeffer2018tampering}, introducing unknown biases at this early stage of data collection~\cite{tromble2017we,OlteanuCDK2019}.

Taking a purely ``big data'' approach can also lead to inappropriate generalisations and conclusions~\cite{lazer2014parable, tufekci2014big}. This is well illustrated, for example, by the range of motivations behind retweeting behaviour including affirmation, sarcasm, disgust and disagreement~\cite{tufekci2014big}. Similarly, in the study of collective action, there are important social interactions that occur offline
~\cite{venturini2018actor}. 
Furthermore, relying solely on observable online behaviours risks overlooking passive consumers, resulting in underestimating the true extent to which social media 
can influence people~\cite{falzon2017representataion}.

OSN APIs provide data by streaming it live or through retrieval services, both of which make use of OSN-specific query syntaxes. Conceptually, therefore, there are two primary collection approaches to consider: 1) focusing on a user or users as seeds (e.g., 
\cite{gruzd2011imagining,morstatter2018alt}) using a snowball strategy~\cite{goodman1961} and 2) using keywords or filter terms (e.g., 
\cite{ratkiewicz2011,morstatter2018alt,woolley2018us,nasim2018real}). 
Focusing on seeds can reveal the flow of information within the communities around the seeds, while a keyword-based collection provides the ebb and flow of conversation related to a topic. 
These approaches can be combined, as exemplified by Morstatter \emph{et al.}~\cite{morstatter2018alt} in their study of the $2017$ German election: an initial keyword-based collection was conducted for eleven days to identify the most active accounts, the usernames of which were then used as keywords in a six week collection.

Once a reasonable dataset is obtained, there may be benefit in stripping junk content included by automated accounts such as bots~\cite{rise2016}. 
The question of whether to remove content from social bots (bots that actively pretend to be human) depends on the research question at hand; because humans are easily fooled by social bots~\cite{
nasim2018real}, their contribution to discussions may be valid (unlike, e.g., that of a sport score announcement bot).

\section{Methodology}

Our initial hypothesis was that if the same collection strategies were used at the same time, then each OSN would provide the same data, regardless of the collection tool used.
Consequently, social networks built from such data using the same methodology should be highly similar, in terms of both network and node level measurements. 
Our methodology consisted of these steps:
\begin{enumerate}
    \item Conduct simultaneous collections on an OSN using the same collection criteria with different tools.
    \item Compare statistics across datasets.
    \item Construct sample social networks from the data collected and compare network-level statistics.
    \item Compare the networks at the node level.
    \item Compare the networks at the cluster level.
\end{enumerate}

\subsection{Scope}

The scope of this work includes datasets obtained via streaming APIs rather than those aggregated through progressively expanding searches, such as follower networks. 
Such collections (especially follower networks) often require data that is prohibitive to obtain, is immediately out of date, and provides no real indication of strength of relationship, as discussed in Section~\ref{sec:online_interactions}. Additionally, in the absence of a domain-focused research question, no particular accounts would make sensible seeds, so we rely on keyword-based collections.

\subsection{Data collection}

Twitter was chosen as the source OSN due to the availability of its data, the fact that the data it provides is highly regular~\cite{JosephLC2014comparison}, and because it has similar interaction primitives to other major OSNs. 
In the interests of comparing a proprietary collection tool with a baseline 
two tools were chosen:

    \textbf{Twarc}\footnote{\url{https://github.com/DocNow/twarc}} is an open source library 
    which wraps 
    Twitter's API, and provided the baseline. 

    \textbf{RAPID} (Real-time Analytics Platform for Interactive Data Mining)~\cite{rapid2017} is a social media collection and data analysis platform for Twitter and Reddit. It enables filtering of OSN live streams, as well as dynamic topic tracking, meaning it can update filter criteria in real time, adding terms popular in recent posts and removing unused ones. 


Both tools facilitate filtering Twitter's Standard live stream\footnote{\url{https://developer.twitter.com/en/docs/tweets/filter-realtime/overview}} with keywords,
providing datasets of tweets as JSON objects.

\subsection{Constructing social networks} \label{sec:net_cons}

A social network is constructed from dyads of pair-wise relations between nodes, which in our case are Twitter accounts. 
The node ties denote intermittent relations between accounts, inferred from observed interactions~\cite{nasim2016inferring, borgatti2009network}. 
Here, 
we consider three social networks 
built from 
interaction types 
common to many OSNs: `mention networks', `reply networks', and `retweet networks' (e.g., retweets are analogous to Facebook shares or Tumblr reposts, and replies analogous to comments on posts on Reddit).
We define a social network, $G$=$(V,E)$, of accounts $u \in V$ linked by directed, weighted edges $(u_i,u_j) \in E$ based on the criteria below.
\textbf{Mention Networks.} Twitter users can \emph{mention} one or more other users in a tweet. In a mention network, an edge $(u_i, u_j)$ exists iff $u_i$ mentions $u_j$ in a tweet, and the weight corresponds to the number of times $u_i$ has mentioned $u_j$.

\textbf{Reply networks.} A tweet can be a reply to one other tweet. In a reply network, an edge $(u_i, u_j)$ exists iff $u_i$ replies to a tweet by $u_j$, and the weight corresponds to the number of replies $u_i$ has made to $u_j$'s tweets. 


\textbf{Retweet networks.} A user can repost or `retweet' another's tweet on their own timeline (visible to their followers). 
Though retweets are not necessarily direct interactions~\cite{RuthsPfeffer2014}, they can be used to determine an account's reach. 
In a retweet network, an edge $(u_i, u_j)$ exists iff $u_i$ retweets a tweet by $u_j$, and its weight corresponds to the number of $u_j$'s tweets $u_i$ has retweeted. 

\begin{figure}[!ht]
    \centering
    \begin{minipage}[t]{\columnwidth}
        \subfloat[Mentions.\label{fig:qanda1_sample_mentions}]{%
            \includegraphics[height=0.16\textheight]{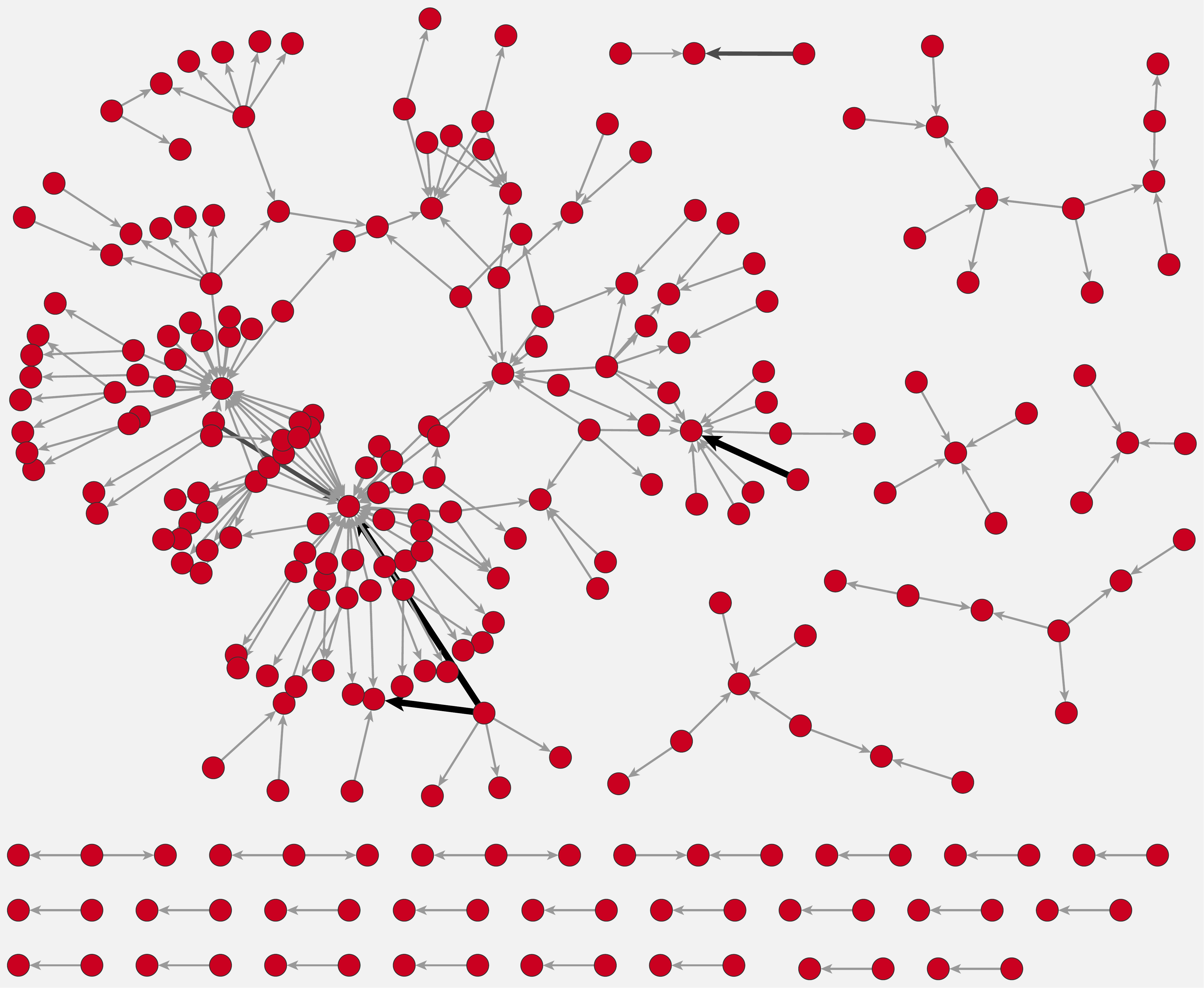}
        }
        \hfill
        \subfloat[Replies.\label{fig:qanda1_sample_replies}]{%
            \includegraphics[height=0.16\textheight]{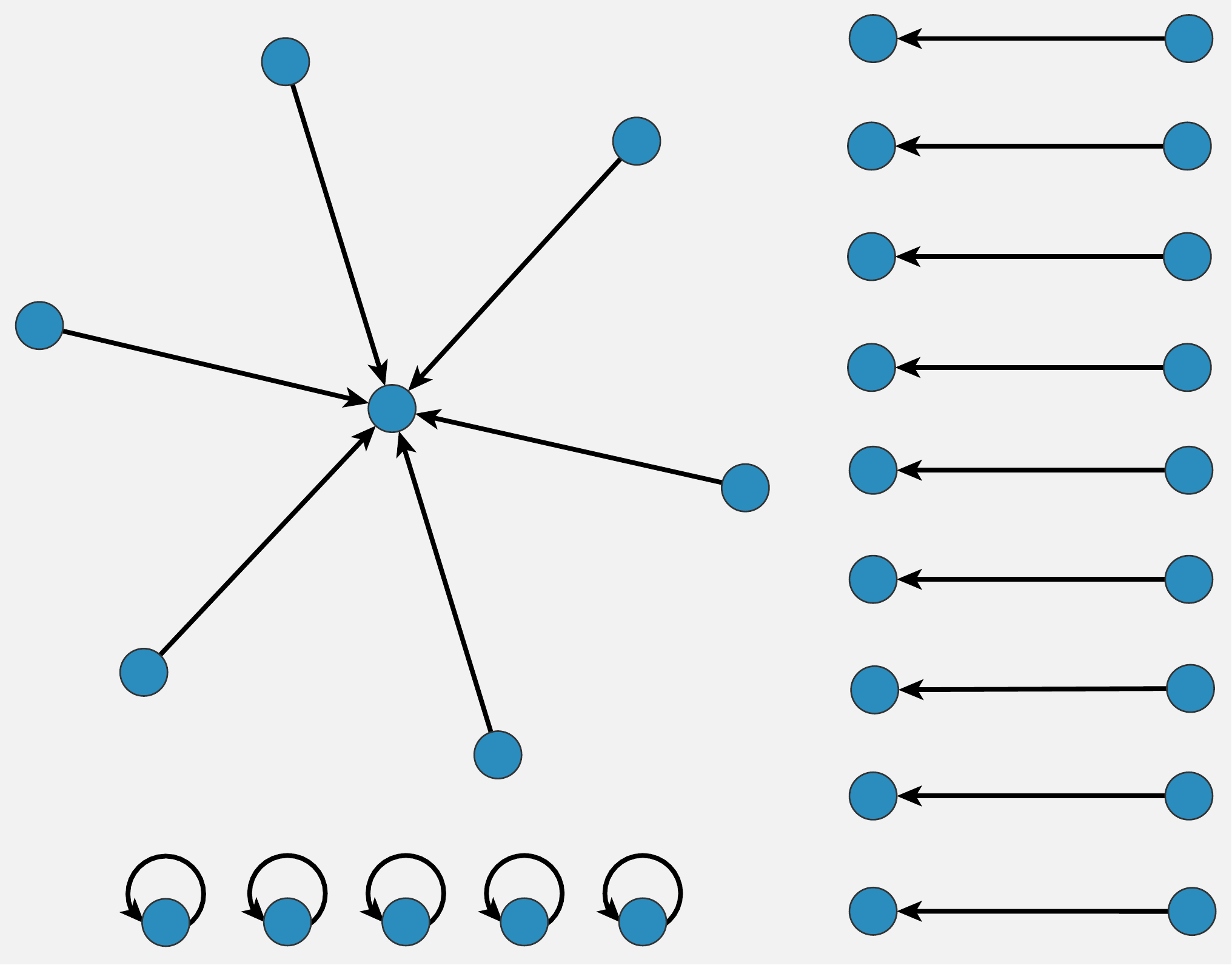}
        }
        \hfill
        \subfloat[Retweets.\label{fig:qanda1_sample_retweets}]{%
            \includegraphics[height=0.16\textheight]{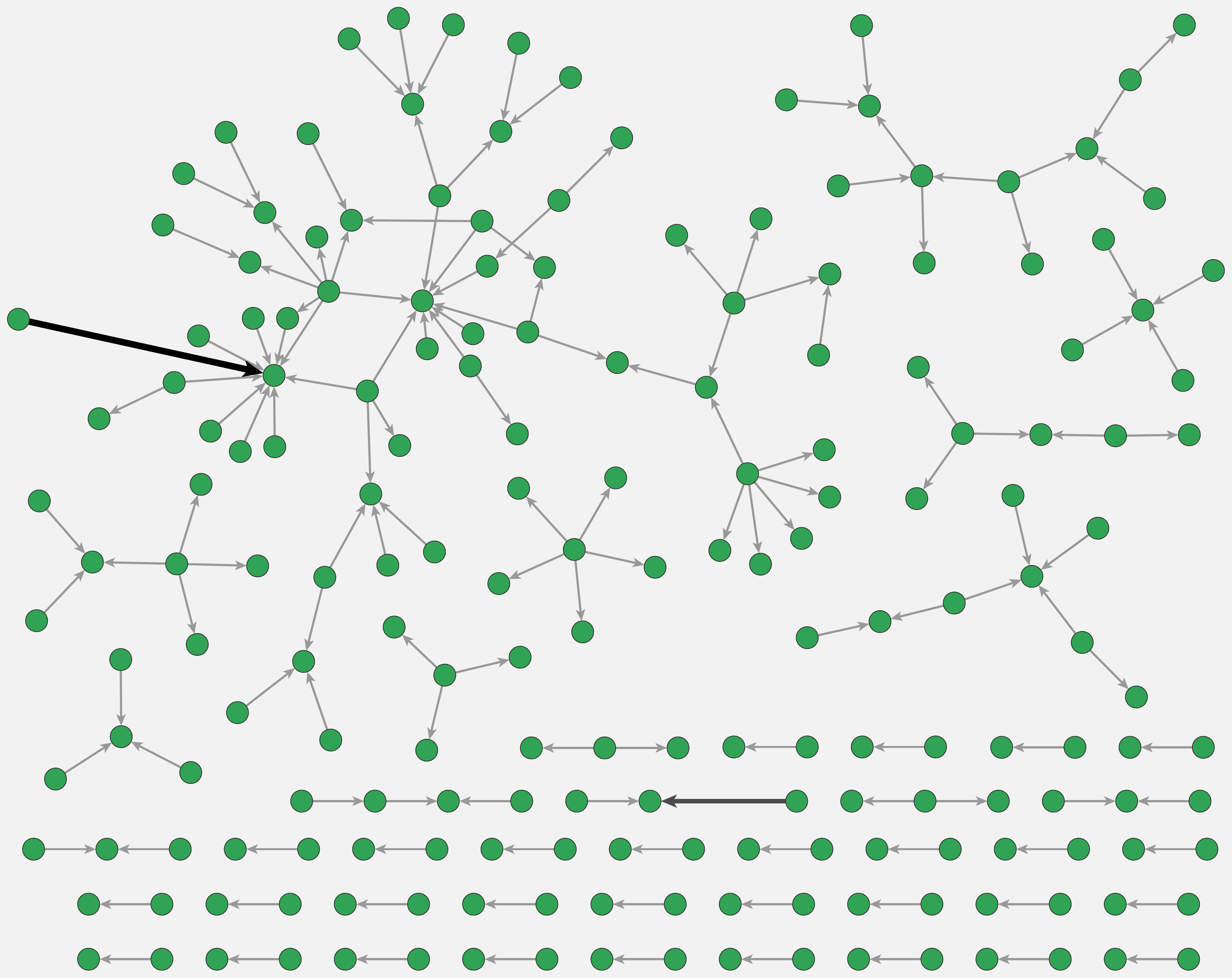}
        }
    \end{minipage}%
    \caption{Sample networks of accounts built from 5 minutes of Twitter data. 
    Nodes may appear in one or more networks, depending on their behaviour during the sampled period. The diagrams were constructed with \textit{visone} (https://visone.info).}
    \label{fig:qanda1_samples}
\end{figure}

Examining networks built from the same dataset, replies (Figure~\ref{fig:qanda1_sample_replies}) are the least common of the three interaction types, and all are dominated by a single large component. Mention networks exhibit relatively high cohesiveness. The similarity between retweets (Figure~\ref{fig:qanda1_sample_retweets}) and mentions (Figure~\ref{fig:qanda1_sample_mentions}) is because the data model of a retweet includes a mention of the retweeted account.

Other network types can be constructed based on direct or inferred relations, including shared use of hashtags or URLs, reciprocation or minimum levels of interaction, or friend/follower connections.
As our focus is on the social relationships implied by direct interactions we will focus only on the above three types of network construction here,
however similar issues will certainly arise in other applications, e.g., narrative analysis~\cite{edwards2018one}.


\subsection{Analyses} \label{sec:analyses}

\subsubsection{Dataset statistics}To compare the parallel datasets we examined the following features:

    \textit{Absolute counts}: accounts, tweets, retweets, quotes, replies, URLs, hashtags, and mentions; and
    
    \textit{Highest counts}: tweeting account, mentioned account, retweeted tweet, replied-to tweet, used hashtags, and used URLs.

Based on these figures, we account for major discrepancies, which can guide post-processing (e.g., spam filtering).

\subsubsection{Network statistics} The following network statistics are used to assess differences in the constructed networks: 
number of nodes, edges, average degree, density, mean edge weight, component count and the size and diameter of the largest, 
Louvain~\cite{blondel2008louvain} cluster count and the size of the largest, reciprocity, transitivity, and maximum k-cores.

\subsubsection{Centrality values} 

Centrality measures offer a way to consider the importance of individual nodes within a network. The centrality measures considered here include: \emph{degree} centrality, indicating how many other nodes one node is directly linked to; \emph{betweenness} centrality, referring to the number of shortest paths in the network that a node is on and thus to what degree the node acts as a bridge between other nodes; \emph{closeness} centrality, which provides a sense of how topologically close a node is to the other nodes in a network; and \emph{eigenvector} centrality, which measures how connected a node is to other highly-connected nodes. 

Only centrality measures for mention and reply networks are considered, as edges in retweet networks are not direct interactions~\cite{RuthsPfeffer2014}.

Given the set of nodes in each corresponding pair of networks is not guaranteed to be identical, it is not possible to directly compare the centrality values of each node, so instead we rank the nodes in each network by the centrality values, take the top $1,000$ from each list, further constrain the lists to only the nodes common to both lists, and then compare the rankings. We initially compare the rankings visually using scatter plots
, where a node's rank in the first and second list is shown on the $x$ and $y$ axes, respectively.
A statistical measure of the similarity of the two rankings (of common nodes) is obtained with  
the Kendall $\tau$ coefficient. A $\tau$ score of $0.4$ is considered to indicate reasonable similarity~\cite{landis1977measurement}. We also calculate Spearman's $\rho$ coefficient as a confirmation measure. 

\subsubsection{Cluster comparison} The final step is to consider the clusters discoverable in the mention, reply and retweet networks and compare their membership. We first compare the distribution of the sizes of the twenty largest Louvain clusters~\cite{blondel2008louvain} visually (chosen because it works well with large and small networks~\cite{yang2016comparative}).

We use the Adjusted Rand index~\cite{HubertArabie1985adjrandindex} to compare membership. 
This considers two networks of the same nodes that have been partitioned into subsets. When considered in pairs, there are nodes that appear in the same subset in both partitions ($a$), and there are (many) pairs of nodes that do not appear in the same subsets in either partition ($b$), and the rest appear in the same subset in one of the partitions but not in the other. Defining the total of possible pairings of the $n$ nodes (\(\frac{n(n-1)}{2}\)) as $c$, the Rand index, $R$, is simply \(R = \frac{a + b}{c}\). The Adjusted Rand index (ARI) corrects for chance and provides a value in the range $[-1,1]$ where $0$ implies that the two partitions are random with respect to one another and~$1$ implies they are identical.

\section{Experiment}


\subsection{Data collection}

To obtain a moderately active portion of activity, we collected data from Twitter's Standard live stream\footnote{All data were collected, stored, processed and analysed according to two ethics protocols 
\#170316 and H-2018-045, approved by the University of Adelaide's 
human research and ethics committee.} relevant to an Australian television panel show that invites its viewers to participate in the discussion live\footnote{The Australian Broadcasting Commission's ``Q\&A'' observes \#QandA.}. 
A particular broadcast in $2018$ was chosen due to the expectation of high levels of activity given the planned discussion topic. 
As a result, the filter keywords used were 
`qanda'\footnote{The `\#' was omitted to catch mentions of `@qanda', the programme's Twitter account.} and two terms 
that identified 
a panel member (available on request). 
We collected two parallel datasets: 

    \textbf{Part 1:} Four hours starting 
    $30$ minutes before the hour-long programme, 
    to allow for contributions from the country's major timezones; and
    
    \textbf{Part 2:} From 6am to 9pm the following day, capturing 
    further related online discussions. 

Twarc acted as the baseline collection as it provides direct access to Twitter's API, 
while 
RAPID was configured to use \emph{co-occurrence keyword expansion}~\cite{rapid2017}, meaning it would progressively add keywords to the original set if they appeared sufficiently frequently (five times in ten minutes).
This expanded dataset was referred to as `RAPID-E' and was filtered back to just the tweets containing the original keywords (and labelled `RAPID') for comparison with the `Twarc' dataset. 
We expected the moderate activity observed would not breach rate limits, and thus RAPID should capture all tweets captured by Twarc. This was not the case (Table~\ref{tab:qanda_collection_stats}).

\begin{table}[t!h]
    \centering
    \caption{Summary dataset statistics.}
    \label{tab:qanda_collection_stats}
    \resizebox{\columnwidth}{!}{%
    \begin{tabular}{@{}llrrrrr@{}}
        \toprule
                       & Dataset & All    & Unique    & Retweets & All      & Unique    \\
                       &         & Tweets & Tweets    &          & Accounts & Accounts  \\
        \midrule
        Part 1         & Twarc   & 27,389 &    11,481 &   14,191 &    7,057 &     2,090 \\
                       \cmidrule{2-7}
        (20:00-00:00)  & RAPID   & 15,930 &        22 &    8,744 &    4,970 &         3 \\
                       & RAPID-E & 17,675 &     1,767 &    9,767 &    5,547 &       527 \\
        \midrule
        Part 2         & Twarc   & 15,490 &     4,089 &   10,988 &    5,799 &     1,128 \\
                       \cmidrule{2-7}
        (06:00-21:00)  & RAPID   & 11,719 &       318 &    8,051 &    4,708 &        37 \\
                       & RAPID-E & 23,583 &    12,180 &   13,679 &    8,854 &      4,007 \\
        \bottomrule
    \end{tabular}
    } 
\end{table}

\subsection{Comparison of collection statistics}

The first striking difference between the datasets was the number of tweets collected and the effect on the number of contributors (Table~\ref{tab:qanda_collection_stats}). 
RAPID collected fewer tweets by fewer accounts, but the datasets were close to subsets of the Twarc datasets. 
Between 26\% and 42\% of the tweets collected by Twarc were missed by RAPID, but the proportion of retweets in each part is similar (52\% and 55\% for Part 1 and 69\% and 71\% for Part 2). 
In both parts, very few accounts appear in only the RAPID collections. 
Discussions with RAPID's developers revealed it dumps tweets that miss the filter terms from the textual parts of tweets (e.g., the body, the author's screen name, and the author's profile description). The extra tweets RAPID collected were relevant and in English\footnote{Sometimes short or obscure filter terms, like `qanda', have meanings in non-target languages.} (based on manual inspection) but posted by different accounts (unique to RAPID-E). Of the tweets that RAPID collected which contained the keywords, they were posted by almost the same accounts as Twarc, but simply did not contain the same tweets.

The benefit of topic tracking via keyword expansion is yet to be strongly evaluated, but this study indicates there are benefits (relevant tweets) as well as costs (missed matching tweets). The rest of this analysis explores how much of a difference it makes with regard to SNA.

\begin{table}[ht]
    \centering
    \caption{Detailed dataset statistics.}
    \label{tab:more_qanda_collection_stats}
    \resizebox{\columnwidth}{!}{%
        \begin{tabular}{@{}l|rr|rr|@{}}
            \toprule
                                               &    \multicolumn{2}{c}{Part 1} &    \multicolumn{2}{c}{Part 2} \\ 
                                               &         RAPID &         Twarc &         RAPID &         Twarc \\
            \midrule
            Tweets                             &        15,930 &        27,389 &        11,719 &        15,490 \\
            \midrule
            Quotes                             &           325 &         1,203 &           498 &         1,232 \\
            Replies                            &         1,446 &         2,067 &         1,715 &         1,731 \\
            Tweets with hashtags               &        10,043 &        15,591 &         3,912 &         3,961 \\
            Tweets with URLs                   &         2,470 &         4,029 &         3,106 &         4,074 \\
            Most prolific account              & Account $a_1$ & Account $a_1$ & Account $a_2$ & Account $a_3$ \\
            Tweets by most prolific account    &           103 &           146 &            57 &            68 \\
            Most retweeted tweet               &   Tweet $t_1$ &   Tweet $t_1$ &   Tweet $t_2$ &   Tweet $t_2$ \\
            Most retweeted tweet count         &           260 &           288 &           385 &           385 \\
            Most replied to tweet              &   Tweet $t_3$ &   Tweet $t_3$ &   Tweet $t_4$ &   Tweet $t_4$ \\
            Most replied to tweet count        &            55 &           121 &            58 &            58 \\
            Tweets with mentions               &        11,314 &        18,253 &        10,472 &        13,514 \\
            Most mentioned account             & Account $a_4$ & Account $a_4$ & Account $a_4$ & Account $a_4$ \\
            Mentions of most mentioned account &         2,883 &         3,853 &         2,753 &         2,752 \\
            Hashtags uses                      &        15,700 &        23,557 &         7,672 &         7,862 \\
            Unique hashtags                    &         1,015 &         1,438 &           960 &         1,082 \\
            Most used hashtag                  &       \#qanda &       \#qanda &       \#qanda &       \#qanda \\
            Uses of most used hashtag          &        10,065 &        15,644 &         2,545 &         2,549 \\
            Next most used hashtag             &      \#auspol &      \#auspol &      \#auspol &      \#auspol \\
            Uses of next most used hashtag     &         1,381 &         2,103 &         1,652 &         1,349 \\
            URLs uses                          &           913 &         1,650 &         1,602 &         2,411 \\
            Unique URLs                        &           399 &           560 &           658 &           790 \\
            Uses of most used URL              &            49 &           128 &            71 &            81 \\
            \bottomrule
        \end{tabular}
    } 
\end{table}

Table~\ref{tab:more_qanda_collection_stats} reveals that although feature counts vary significantly, many of the most common values are the same (e.g., most retweeted tweet, most mentioned account, most used hashtags). 
Many are approximately proportional to corpus size (Twarc is 1.72 and 1.32 times larger than RAPID for Parts 1 and 2, respectively), but with notable exceptions and no apparent pattern. 
Some values are remarkably similar, despite the size of the corpora they arise from being so different. 
For example, Twarc picked up nearly $8,000$ more hashtag uses than RAPID in Part 1, but fewer than $200$ more in Part 2. 
Both Part 1 datasets have the top ten hashtags, in slightly different order. Approximately $5,000$ of the extra hashtag uses are of `\#qanda'. In Part 2, again, the top ten hashtags are nearly the same, but this time the usage counts are similar, except for `\#auspol' being used $22\%$ more often in RAPID ($1,652$ times compared with $1,349$), which would account for the overall difference of $190$ uses when combined with the noise of lesser used hashtags.

\subsection{Comparison of network statistics}


Given the differences in datasets, we expect differences in the derived social networks (Tables \ref{tab:qanda1_graph_stats} and \ref{tab:qanda2_graph_stats})~\cite{HolzmannAK2018}. 
Each network is dominated by a single large component, comprising over 90\% of nodes in the retweet and mention networks, and around 70\% in the reply networks.
The distributions of component sizes appear to follow a power law, 
resulting in 
corresponding high numbers of detected clusters.


\begin{table}[t]
    \centering
    \caption{Part 1 network statistics.}
    \label{tab:qanda1_graph_stats}
    \resizebox{\columnwidth}{!}{%
    \begin{tabular}{@{}l|rr|rr|rr@{}}
        \toprule
                          & \multicolumn{2}{c}{RETWEET} & \multicolumn{2}{c}{MENTION} & \multicolumn{2}{c}{REPLY} \\ 
                          & RAPID       & Twarc         & RAPID       & Twarc         & RAPID       & Twarc       \\
        \midrule 
        Nodes             & 3,234       & 4,426         & 4,535       & 6,119         & 1,184       & 1,490       \\
        Edges             & 7,855       & 12,327        & 13,144      & 19,576        & 1,231       & 1,631       \\
        Average degree    & 2.429       & 2.785         & 2.898       & 3.199         & 1.040       & 1.095       \\
        Density           & 0.001       & 0.001         & 0.001       & 0.001         & 0.001       & 0.001       \\
        Mean edge weight  & 1.113       & 1.151         & 1.268       & 1.300         & 1.175       & 1.267       \\
        Components        & 74          & 95            & 86          & 108           & 164         & 192         \\
        Largest component & 3,061       & 4,115         & 4,326       & 5,819         & 829         & 1,081       \\
        - Diameter        & 12          & 12            & 10          & 11            & 15          & 15          \\
        Clusters          & 93          & 115           & 109         & 134           & 186         & 219         \\
        Largest cluster   & 318         & 540           & 731         & 1,348         & 169         & 229         \\
        Reciprocity       & 0.004       & 0.007         & 0.025       & 0.025         & 0.106       & 0.099       \\
        Transitivity      & 0.026       & 0.034         & 0.065       & 0.063         & 0.024       & 0.021       \\
        Maximum k-core    & 11          & 14            & 13          & 16            & 2           & 3           \\ 
        \bottomrule
    \end{tabular}
    } 
\end{table}

\begin{table}[ht]
    \centering
    \caption{Part 2 network statistics.}
    \label{tab:qanda2_graph_stats}
    \resizebox{\columnwidth}{!}{%
    \begin{tabular}{@{}l|rr|rr|rr@{}}
        \toprule
                          & \multicolumn{2}{c}{RETWEET} & \multicolumn{2}{c}{MENTION} & \multicolumn{2}{c}{REPLY} \\ 
                          & RAPID       & Twarc         & RAPID       & Twarc         & RAPID       & Twarc       \\
        \midrule 
        Nodes             & 3,594       & 4,591         & 5,198       & 6,205         & 1,492       & 1,507       \\
        Edges             & 7,344       & 10,110        & 14,802      & 18,184        & 1,560       & 1,576       \\
        Average degree    & 2.043       & 2.202         & 2.848       & 2.931         & 1.046       & 1.046       \\
        Density           & 0.001       & 0.000         & 0.001       & 0.000         & 0.001       & 0.001       \\
        Mean edge weight  & 1.096       & 1.087         & 1.245       & 1.222         & 1.099       & 1.098       \\
        Components        & 118         & 176           & 123         & 179           & 196         & 201         \\
        Largest component & 3,308       & 4,085         & 4,854       & 5,612         & 1,073       & 1,080       \\
        - Diameter        & 12          & 11            & 10          & 10            & 15          & 15          \\
        Clusters          & 138         & 197           & 158         & 210           & 221         & 226         \\
        Largest cluster   & 471         & 727           & 1,090       & 1,513         & 122         & 123         \\
        Reciprocity       & 0.004       & 0.004         & 0.024       & 0.025         & 0.072       & 0.071       \\
        Transitivity      & 0.027       & 0.026         & 0.084       & 0.079         & 0.016       & 0.016       \\
        Maxmium k-core    & 9           & 10            & 11          & 14            & 3           & 3           \\ 
        \bottomrule
    \end{tabular}
    } 
\end{table}

Structural statistics like density, diameter (of the largest component in disconnected networks), reciprocity and transitivity may offer insight into social behaviours such as influence and information gathering. 
The high component counts in all networks lead to low densities and correspondingly low transitivities, as the potential number of triads is limited by the connectivity of nodes. 
That said, the largest components were consistently larger in the Twarc datasets, but the diameters of the corresponding largest components from each dataset were remarkably similar, implying that the extra nodes and edges were in the components' centres rather than on the periphery. 
This increase in internal structures 
improves connectivity and therefore the number of nodes to which any one node could pass information (and therefore influence) or, at least, reduces the length of paths between nodes so information can pass more quickly. 
The similarities in transitivity imply the increase may not be significant, however, with networks of these sizes. Reciprocity values may provide insight into information gathering, which often relies on patterns of to-and-fro communication as a person asks a question and others respond. Interestingly, the only significant difference in reciprocity is in the Part 1 retweet networks, with the Twarc dataset having a reciprocity nearly double that of the RAPID dataset (though still small). The Twarc dataset includes 60\% more retweets than the corresponding RAPID dataset and 40\% more accounts (Table~\ref{tab:qanda_collection_stats}), 
which may account for the discrepancy. 
Given the network sizes, the reciprocity values indicate low degrees of 
conversation, mostly in the reply networks. 
Interestingly, mean edge weights are very low ($1.3$ at most), implying that most interactions between accounts in all networks happen only once, despite these being corpora of issue-based discussions.

Next we look at two major categories of network analysis: \emph{indexing}, for the computation of node-level properties, such as centrality; and \emph{grouping}, for the computation of specific groups of nodes, such as clustering.

\subsection{Comparison of centralities}

Centrality measures can tell us about the influence an individual has over their neighbourhood, though the timing of interactions should ideally be taken into account to get a better understanding of their dynamic aspects (e.g., \cite{falzon2018embedding}). 
If networks are constructed from partial data, network-level metrics (e.g., radius, shortest paths, cluster detection) and neighbourhood-aware measures (e.g., eigenvector and Katz centrality) may vary and not be meaningful~\cite{HolzmannAK2018}.

\begin{figure}[!ht]
    \centering
    \begin{minipage}[t]{0.99\columnwidth}
        \subfloat[Part 1.\label{fig:qanda1_centrality_ranking_comparisons}]{
            \includegraphics[width=\columnwidth]{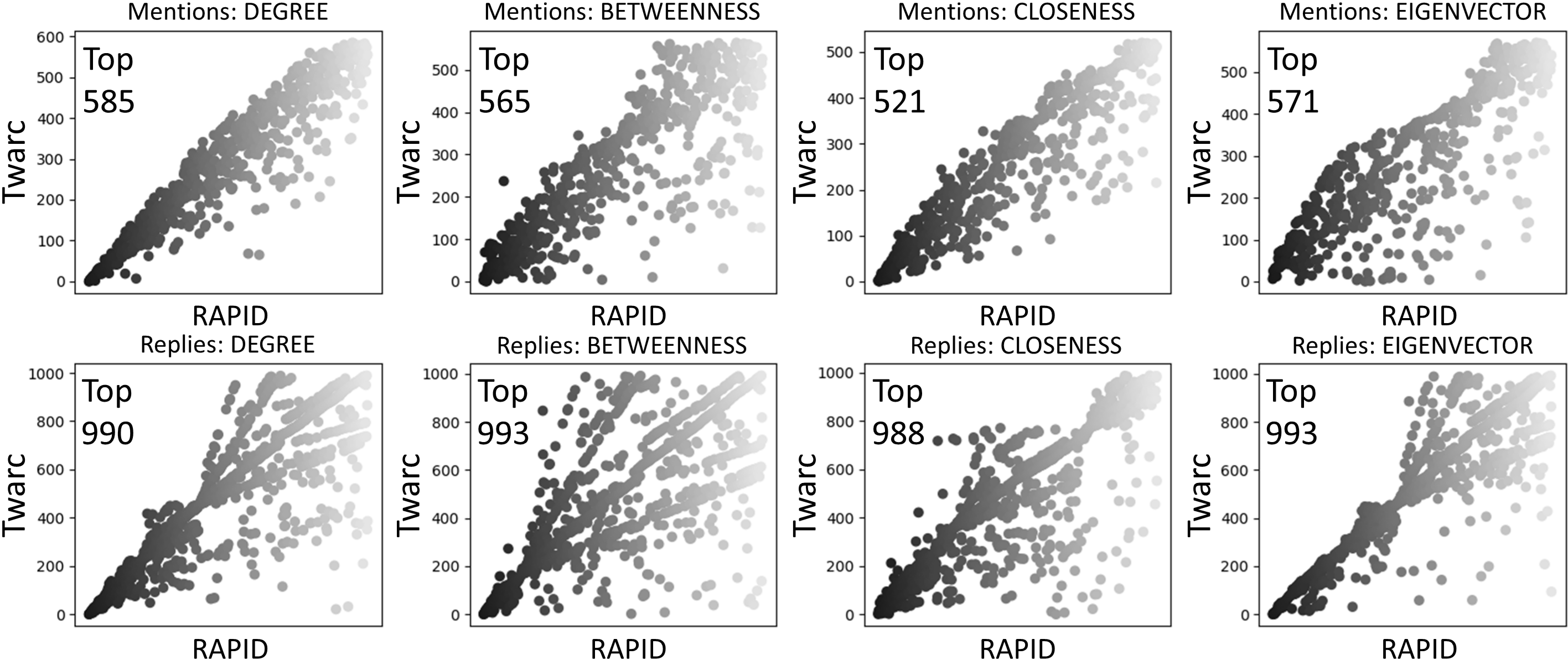}
        }
        \hfill
        \subfloat[Part 2. 
        \label{fig:qanda2_centrality_ranking_comparisons}]{
            \includegraphics[width=\columnwidth]{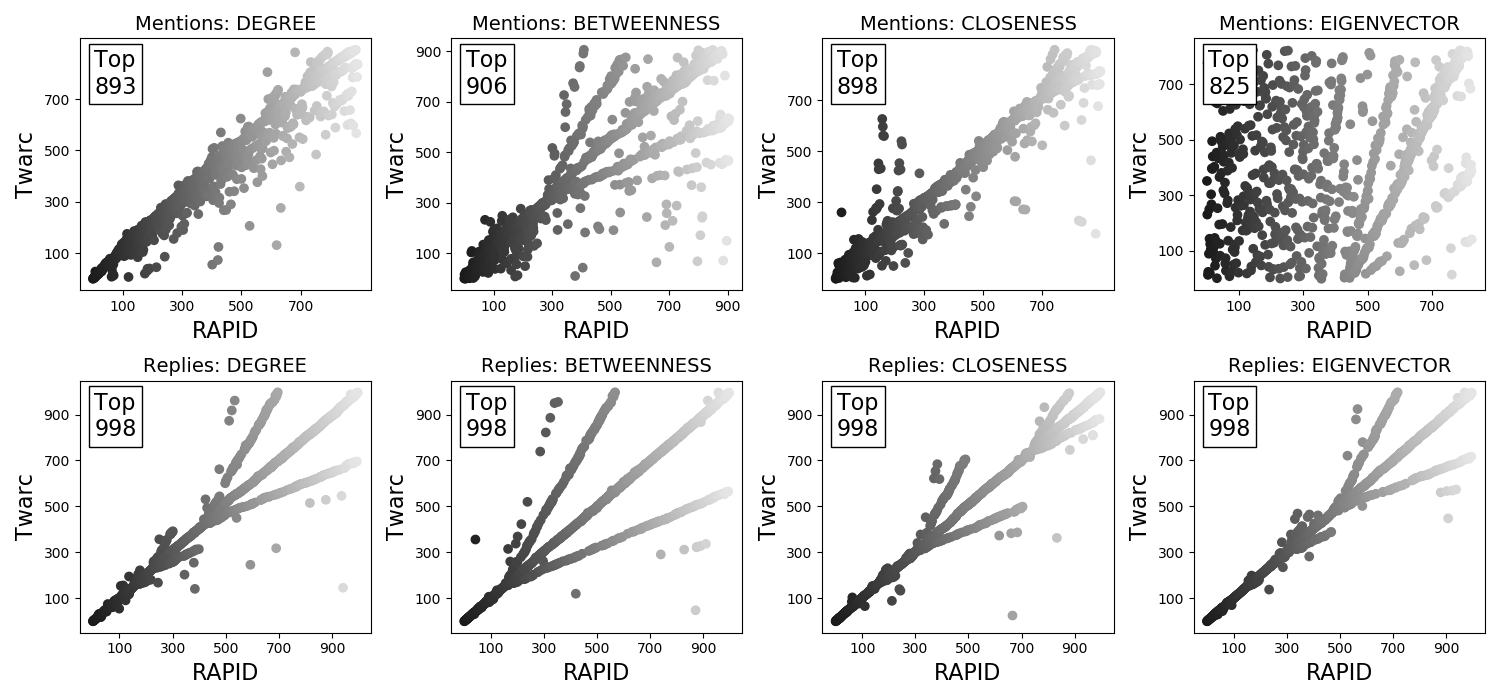}
        }
    \end{minipage}
    \caption{Centrality ranking comparison scatter plots of the mention and reply networks built from the Q\&A Parts 1 and 2 datasets. In each plot, each point represents a node's ranking in the RAPID and Twarc lists of centralities (common nodes amongst the top $1,000$ of each list). The number of nodes appearing in both lists is inset. Point darkness indicates rank on the $x$ axis (darker = higher).}
    \label{fig:qanda_centrality_ranking_comparisons}
\end{figure}

We compare centralities of corresponding networks using scatter plots of node rankings, as per Section~\ref{sec:analyses} (Figure~\ref{fig:qanda_centrality_ranking_comparisons}). 
The symmetrical structures come from corresponding shifts in order: if an item appears higher in one list, then it displaces another, leading to the evident fork-like patterns. 
There is considerable variation in most centrality rankings for both mention and reply networks in Part 1 (Figure \ref{fig:qanda1_centrality_ranking_comparisons}) but much less in Part 2 (Figure \ref{fig:qanda2_centrality_ranking_comparisons}). 
Furthermore, the relatively few common nodes in Part 1's Twarc mention networks ($521$ to $585$) and greater edge count (Tables~\ref{tab:qanda1_graph_stats}) could indicate that the extra edges significantly affect the node rankings. 
However, Part 2's Twarc mentions networks also had many more edges, but many more nodes in common (approximately $900$).
Thus it must have been how the mentions were distributed in the datasets that differed, rather than simply their number.
It is not clear that Part 1's four hour duration (\emph{cf.} Part 2's $15$ hours) explains this. 
Instead, if we look at the $11,480$ tweets unique to Twarc in Part~1 (\emph{cf.} fewer than $4,000$ are unique to Twarc in Part 2, Table~\ref{tab:qanda_collection_stats}), 
only $622$ are replies, whereas $6,915$ include mentions. 
There are also $34\%$ more unique accounts in the Part~1 Twarc dataset, but only $19\%$ more in the Part 2 Twarc dataset (Table~\ref{tab:qanda_collection_stats}). 
Each mention refers to one of these accounts and forms an extra edge in the mention network, thus altering the network's structure and the centrality values of many of its nodes; this is likely where the variation in rankings originates.

\begin{figure}[!ht]
    \centering
    \begin{minipage}[t]{\columnwidth}
        \subfloat[Part 1.\label{fig:qanda1_centrality_ranking_comparisons_tau_rho}]{
            \includegraphics[width=0.99\textwidth]{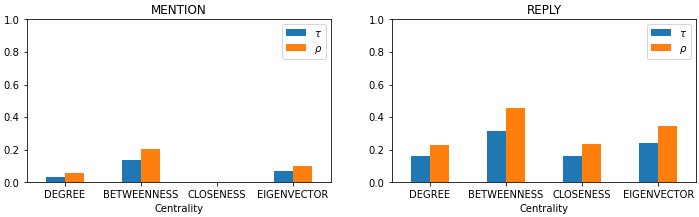}
        }
        \\ 
        \subfloat[Part 2.\label{fig:qanda2_centrality_ranking_comparisons_tau_rho}]{
            \includegraphics[width=0.99\textwidth]{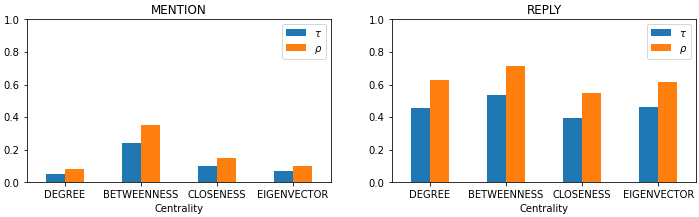}
        }
    \end{minipage}
    \caption{Centrality ranking comparisons using Kendall $\tau$ and Spearman's $\rho$ coefficients
    for corresponding mention and reply networks.}
    \label{fig:qanda_centrality_ranking_comparisons_tau_rho}
\end{figure}


The Kendall $\tau$ and Spearman's $\rho$ coefficients were calculated comparing the corresponding lists of nodes, each pair ranked by one of the four centrality measures 
(Figure~\ref{fig:qanda_centrality_ranking_comparisons_tau_rho}). 
Although somewhat proportional, it is notable how different the coefficient values are, especially in Part~2. 
While Twarc produced more tweets than RAPID (Table~\ref{tab:qanda_collection_stats})
, and more unique accounts, the corresponding mention and reply node counts are not significantly higher (Tables~\ref{tab:qanda1_graph_stats} and~\ref{tab:qanda2_graph_stats}). 
In fact, the node counts in the Part~1 reply networks are correspondingly lower than in the Part~2 reply networks, 
even though both Part~2 datasets were smaller.
Edge counts in the mention networks were very different (Twarc had many more) but were quite similar in the reply networks.

The biggest variation was in the mention networks from Part 1 (Figure~\ref{fig:qanda1_centrality_ranking_comparisons} and Table~\ref{tab:qanda1_graph_stats}), 
due to the large number of extra mentions from Twarc. It is notable that the Kendall's $\tau$ was low for all mention networks (Figure~\ref{fig:qanda_centrality_ranking_comparisons_tau_rho}), especially for degree and closeness centrality. 
It is worth noting the minor differences in the degree and immediate neighbours of nodes impacts degree and closeness centralities significantly, and, correspondingly, their relative rankings. 
In contrast, rankings for betweenness and eigenvector centrality, which rely more on global network structure, remained relatively stable.

\subsection{Comparison of clusters}

\begin{figure}[!ht]
    \begin{minipage}[t]{\columnwidth}
        \subfloat[Part 1.\label{fig:qanda1_top20_clusters}]{
            \includegraphics[width=0.47\textwidth]{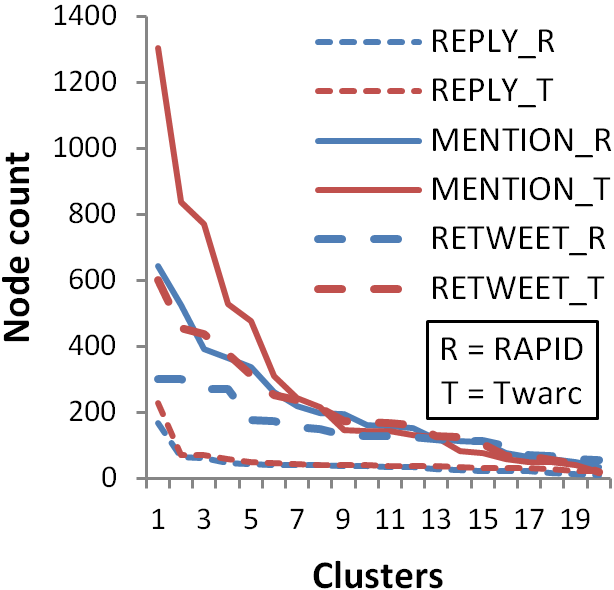}
        }
        \hfill
        \subfloat[Part 2.\label{fig:qanda2_top20_clusters}]{
            \includegraphics[width=0.47\textwidth]{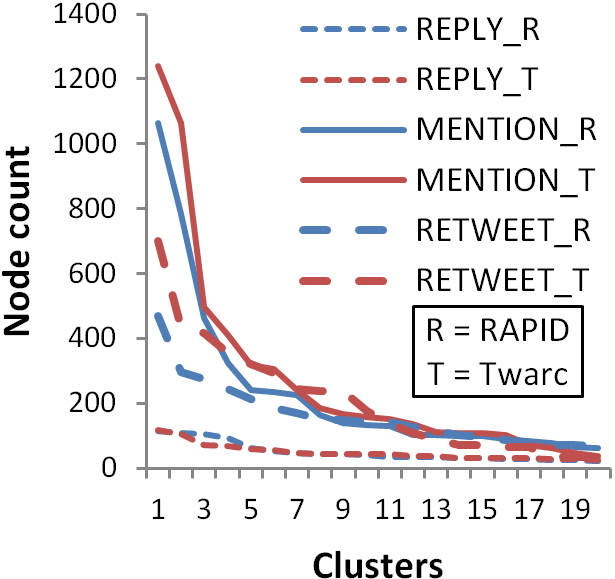}
        }
    \end{minipage}
    \caption{The largest retweet, mention and reply clusters.}
    \label{fig:qanda_top20_clusters}
\end{figure}

We finally compare the networks via largest clusters 
(Figure~\ref{fig:qanda_top20_clusters}).  
The reply network clusters are relatively similar, 
and the largest mention and reply clusters differ the most.
The ARI scores 
confirm that the reply clusters were most similar for Parts~1 and~2 ($0.738$ and $0.756$, respectively), possibly due to the small size of the reply networks. The mention and retweet clusters for Part~2 were more similar than those of Part~1 ($0.437$ and $0.468$ compared to $0.320$ and $0.350$), possibly due to the longer collection period. In Part~1, there is a chance the networks are different due to RAPID's expansion strategy. 
Changes to filter keywords may have collected  
posts of other vocal accounts not using the original keywords, at the cost of the posts 
which did.



\section{Conclusion}






RAPID and Twarc provided very different views of the Q\&A Twitter discussion. 
This manifested as variations in collection statistics, and network-level and node-level statistics for various 
social 
networks built from the collected data. 
The extra tweets collected by Twarc appear to have resulted in more connections in the largest components. This may affect analysis of diffusion, as reachability correspondingly increases. 
Deeper study of reply content is required to inform discussion patterns. 

\color{black}

How reliable social media can be as a source for research without deep knowledge of the effects of collection tools on analyses is an open question. 
If a tool \emph{adds value} through analytics, what is the nature of the effect? 
This paper provides a methodology to explore those effects.

We recommend the following to those using OSN data: 
\begin{itemize}
    \item \vspace{-0.5em}
    Be aware of tool biases and their effects.
    \item Take care to specify filtering conditions with keywords that capture relevant data and avoid irrelevant data. Beware of short terms and ones that are meaningful in non-target languages.

    
    

    \item Check the integrity of data. We observed gaps in another case studies due to connection failures as well as the appearance of duplicate tweets, identical in data and metadata.
    \vspace{-0.5em}
\end{itemize}

Key for future social media research is to develop processes for repeatable analysis. Currently, OSN terms and conditions often hamper researchers exchanging datasets, so new techniques cannot easily be evaluated on the same data, raising questions of fair comparison.

Finally: Does it matter? As long as a researcher makes clear how they conducted a collection and using what tools and configuration, does it not still result in an analysis of behaviour that occurred online?
Answer: It very much depends on the conclusions being drawn. Yes, the collection represents real activity that occurred, but it may not be complete, and conclusions drawn from it may be unintentionally misinformed and lacking in nuance. 
Variations in collections may affect centrality analyses and mislead influential account identification, which, depending on who uses the results of these analyses, may waste effort attending to the wrong topics and/or accounts. 
A firm understanding of the data and how it was obtained is therefore vital.

\bibliographystyle{IEEEtran}
\bibliography{jsocnet}

\end{document}